\documentclass[journal=apchd5,manuscript=article]{achemso}

\usepackage{multirow}

\newcommand*{\mynote}[1]{{
\renewcommand{\baselinestretch}{1.0}\selectfont
\raggedright\noindent\rule{0mm}{6mm}\texttt{$\star$ \textsl{#1}}
\vspace{1mm}\par
}}

\author{Andrzej Szczepkowicz}
\email{andrzej.szczepkowicz@uwr.edu.pl}
\phone{+48 713759304}
\author{Dmytro Konakhovych}
\author{Damian \'{S}nie\.{z}ek}
\affiliation[University of Wroclaw]
{Institute of Experimental Physics, University of Wroclaw, Plac Maxa Borna 9, 50-204 Wroclaw, Poland}
\author{Dylan S. Black}
\affiliation{Department of Electrical Engineering, Stanford University, 350 Serra Mall, Stanford, California 94305-9505, USA}
\author{R. Joel England}
\affiliation{SLAC National Accelerator Laboratory, 2575 Sand Hill Road, Menlo Park, California 94025, USA}
\author{Yen-Chieh Huang}
\affiliation{Institute of Photonics Technologies, National Tsing Hua University, Hsinchu 30013, Taiwan}
\author{Levi Sch\"{a}chter}
\affiliation{Technion--Israel Institute of Technology, Haifa 32000, Israel}

\title[Efficiency of gratings\ldots]
  {Efficiency of gratings for silica fiber-coupled internal Smith-Purcell radiation and Cherenkov diffraction radiation -- a quantitative numerical study}

\keywords{Smith-Purcell radiation, Cherenkov radiation, diffraction radiation, dielectric grating, optical fiber, integrated photonics, particle beam monitor}

\begin{document}

\fbox{\fbox{\fbox{\parbox{0.8\textwidth}{
\mynote{\textbf{Compiled \date{\today}.}}
}}}}\newpage

\begin{tocentry}

\includegraphics[width=8cm]{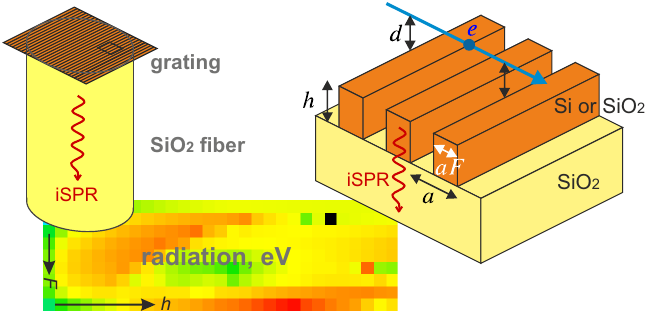}

\end{tocentry}

\begin{abstract}
We propose a setup for measuring visible and near-visible internal Smith-Purcell radiation and Cherenkov Diffraction Radiation, based on silica and silicon, and perform quantitative numerical analysis of its radiation efficiency.
We calculate the total radiated energy per electron and the spectral distribution of different radiation orders, taking into account material dispersion and absorption. For an optimized silica grating of 200 micrometer length, the total radiated energy reaches 2~eV per electron for 2 MeV electrons.
Above the Cherenkov threshold, in most cases the energy of Cherenkov diffraction radiation is several times higher than the energy of internal Smith-Purcell radiation, but for some geometries and frequency ranges first or second order radiation may dominate over Cherenkov radiation (zeroth order). Radiation up to the 4th order is detected in the simulation. The spectrum of Cherenkov radiation is highly resonant, with local minima for frequencies where maxima of the other radiation orders occur. The spectrum of Cherenkov radiation takes a frequency comb-like shape for a uniform layer of Si on SiO2 substrate, due to a Fabry-Perot effect occurring for the evanescent field of the moving electron.
The proposed setup could become a prototype of a non-invasive particle beam monitor for both conventional and laser particle accelerators.
\end{abstract}

\vspace{10mm}\noindent\begin{minipage}{13cm}
   \renewcommand{\baselinestretch}{1}
   \tableofcontents 
\end{minipage}\vspace{8mm}

\section{Introduction}

\begin{figure}
\includegraphics{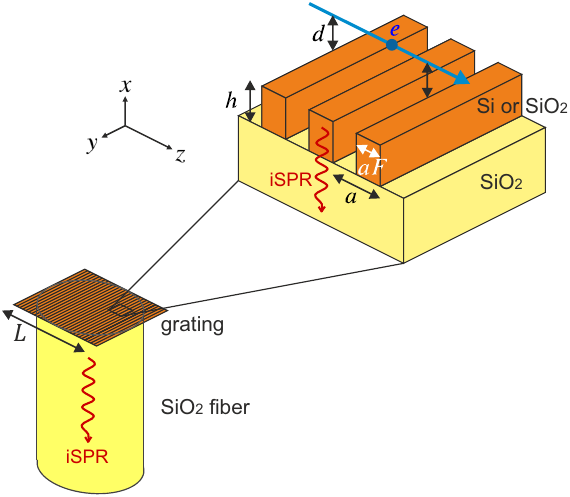}
\caption{Proposed configuration for generating and coupling ISPR and CDR: a rectangular grating attached to a multimode optical fiber.}
\label{fig-geometry}
\end{figure}

The interaction of electron beams with photonic structures is an active area of research, with potential applications in novel on-chip devices like tunable light sources, miniaturized particle accelerators and particle detectors\cite{2023-Roques-Carmes-Kooi, 2022-Shiloh-Schonenberger, 2012-Soong-Byer}.
The emission of internal Smith--Purcell Radiation (iSPR) and Cherenkov Diffraction Radiation (CDR) are two examples of this class of phenomena\cite{2021-Konakhovych-Sniezek,2022-Konakhovych-Sniezek, 2000-Takahashi-Shibata, 2020-Curcio-Bergamashi}. Both are initiated by electrons travelling in vacuum in proximity of a dielectric region and both generate light which propagates inside the dielectric. The iSPR is emitted when the surface of the dielectric is periodically corrugated (``grating''), while CDR is emitted when the particle in vacuum is faster than light in the dielectric. Obviously both requirements may be satisfied at the same time, resulting in a complex radiation spectrum and requiring decomposition into radiation orders to differentiate between iSPR and CDR.
While conventional Smith--Purcell Radiation (SPR) into the vacuum is relatively well studied, so far little is known about its internal variant, iSPR, and its interplay with CDR\cite{2021-Konakhovych-Sniezek,2022-Konakhovych-Sniezek}.
In this work we use a numerical model to study iSPR and CDR with the following new goals:
\begin{enumerate}
    \item\label{goal-easy} Propose and optimize a simple setup for generating and collecting iSPR and CDR which is easy to fabricate with established photonic semiconductor nanofabrication techniques.
    \item Compute the radiation efficiency of the above setup in concrete units: electronvolts per one electron.
    \item Decompose the radiation field obtained in the numerical model into radiation orders, to distinguish between iSPR and CDR and to show energy spectra separately for every radiation mode.
    \item Perform the study in a wide frequency range (visible + near infrared + near ultraviolet), taking material absorption and dispersion into account.
\end{enumerate}
With goal (\ref{goal-easy}) in mind, we propose to use dielectric materials popular in photonics industry: silica, or a combination of silicon and silica, and a conventional multimode silica optical fiber for efficient direct radiation coupling, without free space propagation of light. The envisioned setup is shown in Fig.~\ref{fig-geometry}. This could be a prototype of new, efficient, non-invasive particle beam monitor for both conventional and miniaturised laser particle accelerators \cite{2020-Curcio-Bergamashi,2022-Shiloh-Schonenberger}.

\section{Methods}

\subsection{Radiation frequency}

The frequency of iSPR is given by 
\cite{2012-Ren-Deng,2015-Ponomarenko-Lekomtsev,2021-Konakhovych-Sniezek}
\begin{equation}\label{eq-iSPR-freq}
f=
\frac{c}{a}
\cdot
\frac{m}{\beta^{-1}-n(f)\cos\theta},
\end{equation}
where $c$ denotes speed of light in vacuum, $a$ is the grating period, $m$ is radiation order, $\beta$ is electron velocity in units of $c$, $n(f)$ is the frequency-dependent refractive index of the dielectric grating (or of the grating support in case of gratings composed of two materials), and $\theta$ is the angle 
between direction of the electron and direction of radiation. Note that Eq.~\ref{eq-iSPR-freq} becomes the ordinary vacuum SPR formula when the refractive index of the substrate $n(f)$ is replaced by the refractive index of vacuum. 

To emphasize the connection between iSPR and CDR, we transform Eq.~(\ref{eq-iSPR-freq}) into an alternative form:
\begin{equation}\label{eq-iSPRCDR-freq}
\frac{1}{\beta}-n(f)\cos\theta=\frac{m c}{a f}.
\end{equation}
This form is more general, applicable both for iSPR ($m\neq0$) and CDR ($m=0$). In the latter case (\ref{eq-iSPRCDR-freq}) becomes the equation for the Cherenkov angle. In this way iSPR and CDR can be treated as different radiation orders $m$ of the same radiation field. Another way to see this is described in Sect.~\ref{sect-methods-decomposition}, where the two radiation orders are described by adjacent terms in the Fourier expansion of the full radiation field.

In this work we consider radiation in the 200--1200 THz frequency range (250--1500 nm wavelength range). We take material dispersion and absorption into account\cite{refractiveindex-info}. In the considered frequency range, the refractive index $n(f)$ is in the range 1.44--1.51 for silica and 3.48--6.82 for silicon, while the extinction coefficient $k(f)$ is negligible for silica and 0--5.31 for silicon. The frequency $f$ appears on both sides of Eq.~(\ref{eq-iSPR-freq}), so the equation has to be solved iteratively for $f$, for each set of parameters $m,\theta$. In particular, setting $\theta=0^{\circ}$ and $\theta=180^{\circ}$ gives the equation for the frequency range of each mode -- the solutions are given in Tables \ref{table-f-ranges-30} and \ref{table-f-ranges-2000}. 
When the angular collection aperture for radiation is limited, the effective frequency range of each collected radiation mode is reduced. This is shown in the last two columns of Tables \ref{table-f-ranges-30} and \ref{table-f-ranges-2000} for two values of numerical aperture (NA), as explained in Sect.~\ref{sect-fiber}.

{\renewcommand{\arraystretch}{1.2}
\begin{table}
  \caption{Frequency ranges of iSPR for 30 keV electrons.}
  \label{table-f-ranges-30}
  \begin{minipage}{\textwidth}
  \begin{tabular}{|l|l|l|l|l|l|l|}
    \hline
    Radiation   & \multicolumn{6}{l|}{Frequency limits (minimum, maximum) [THz]}\\
    \cline{2-7}
    order $m$        &   \multicolumn{2}{l|}{Full (all angles)\ \ \ }  & \multicolumn{2}{l|}{Limited to NA=1.1}  & \multicolumn{2}{l|}{Limited to NA=0.22} \\
    \hline
         & Min&Max & Min&Max & Min&Max\\
    \cline{2-7}
    $+1$ & 277.9\footnote{410 THz in the normal direction.}&794.4  & 305.0&632.1 & 382.6&442.1 \\
    $+2$ & 554.5&$>$1200\footnote{Frequency above the scan range of the numerical study is denoted by ``$>$1200''.} & 607.4&$>$1200 & 765.1&884.3    \\
    $+3$ & 829.0&$>$1200 & 905.9&$>$1200 & 1147.7&$>$1200 \\
    $+4$ & 1100.0&$>$1200 & 1197.4&$>$1200 & $>$1200&$>$1200\\
    \hline
  \end{tabular}
  \end{minipage}  
\end{table}}

{\renewcommand{\arraystretch}{1.2}
\begin{table}
  \caption{Frequency ranges of iSPR and CDR orders for 2 MeV electrons.}
  \label{table-f-ranges-2000}
   \begin{minipage}{\textwidth}
   \begin{tabular}{|l|l|l|l|l|l|l|}
    \hline
    Radiation   & \multicolumn{6}{l|}{Frequency limits (minimum, maximum) [THz]}\\
    \cline{2-7}
    order $m$        &  \multicolumn{2}{l|}{Full (all angles)\ \ \ } & \multicolumn{2}{l|}{Limited to NA=1.1} & \multicolumn{2}{l|}{Limited to NA=0.22} \\
    \cline{2-7}
     & Min&Max & Min&Max & Min&Max\\
    \hline
    $-2$  & $>$1200\footnote{Frequency above the scan range of the numerical study is denoted by ``$>$1200''.} & $>$1200  & ~~---\footnote{Not captured within the aperture.} &  ~~--- & ~~--- &  ~~--- \\
    $-1$  & 911.5&$>$1200 & ~~--- & ~~--- & ~~--- & ~~--- \\
    $\phantom{-}0$    & 0&$>$1200 & 118.6& $>$1200 & ~~--- & ~~---  \\
    $+1$ & 170.0\footnote{410 THz in the normal direction.}&$>$1200  & 202.8&$>$1200 & 337.3&522.5 \\
    $+2$ & 338.6&$>$1200 & 403.1&$>$1200 &   674.6&1045.0 \\
    $+3$ & 506.5&$>$1200 & 601.4&$>$1200 & 1011.9&$>$1200 \\
    $+4$ & 673.3&$>$1200 & 796.5&$>$1200 & $>$1200&$>$1200 \\
    $+5$ & 838.4&$>$1200 & 986.8&$>$1200 & $>$1200&$>$1200  \\
    $+6$ & 1001.2&$>$1200 & 1171.1&$>$1200 & $>$1200&$>$1200  \\
    \hline
  \end{tabular}
   \end{minipage}
\end{table}}

In contrast to conventional SPR, for iSPR negative radiation orders are possible ($m<0$ in Eq.~\ref{eq-iSPR-freq}) \cite{2015-Ponomarenko-Lekomtsev}, but only for velocities above the Cherenkov threshold, $\beta>n^{-1}$; of the two electron kinetic energies studied here, 30~keV and 2~MeV, only the latter meets this requirement. For negative orders, the radiation angle $\theta$ is smaller than the Cherenkov angle, and the smallest frequency is for $\theta=0$. In this study, negative orders give negligible contribution to radiation, but for completeness they are mentioned in Table~\ref{table-f-ranges-2000}.

\subsection{The gratings}

In this work we assume gratings of simple rectangular geometry. The grating period $a$ is chosen such that the frequency of first-order iSPR radiation normal to the grating is 410~THz (731~nm). Equation (\ref{eq-iSPR-freq}) then implies $a=240$~nm for 30~keV electrons and $a=716$~nm for 2~MeV electrons. In both cases we assume the total length of the grating to be $L=200~\mu$m, matching an optical fiber with corresponding diameter $L$.

For grating material we consider two materials popular in photonics industry: (1) pure SiO$_2$ grating, and (2) Si bars on SiO$_2$ support (Si/SiO$_2$), see Fig.~\ref{fig-geometry}. We assume that both types of gratings will be bonded directly to a flat cross-section of a SiO$_2$ fiber.

\subsection{\label{sect-fiber}Fiber coupling of radiation}

We assume that iSPR propagating inside SiO2 grating support is directly captured by a multimode SiO2 fiber with a large core diameter $L=200~\mu$m (e.g.\ Thorlabs FG200UEP). The diameter is much larger than radiation wavelength, so modal characteristics of the fiber is not important, and iSPR generated in the grating is guided by the fiber as long as the radiation angle is within the numerical aperture (NA) limit. To obtain frequency limits for a given NA of the fiber, we replace
$n(f)\cos\theta$ in Eq.~\ref{eq-iSPR-freq} with $\pm$NA. The result is shown in Tables~\ref{table-f-ranges-30} and \ref{table-f-ranges-2000}. One experimental option is to use the nominal numerical aperture of the fiber (NA = 0.22 for Thorlabs FG200UEP). Another option is to remove the plastic coating of the fiber and rely on light guiding by total internal reflection at the vacuum interface. In the latter case we obtain NA = $\sqrt{n^2-1} \approx 1.1$ and a wider frequency range can be captured by the fiber, as shown in Tables \ref{table-f-ranges-30} and \ref{table-f-ranges-2000}.

\subsection{Numerical simulation method}

All radiation results in this paper come from a frequency-domain numerical simulation\cite{2020-Szczepkowicz-Schachter-England} for an infinite grating, and are subsequently scaled to assumed grating length of $L=200\ \mu$m. Because the grating is invariant in the $y$ direction (Fig.~\ref{fig-geometry}), all essential features of radiation are captured using a 2D model. With a proper choice of 2D surface current density\cite{2020-Szczepkowicz-Schachter-England} we obtain quantitative data valid for a real 3D geometry which is accurate within $\pm40\%$\cite{2020-Szczepkowicz-Schachter-England,2021-Konakhovych-Sniezek,2022-Konakhovych-Sniezek}.

\subsection{Significance of the impact parameter}\label{sect-impact-par}

Energy radiated from the electron at frequency $f$ depends on its impact parameter $d$ (distance from the surface). Both iSPR and CDR are special cases of diffraction radiation, and radiated energy $W$ is proportional to the exponential factor\cite{1955-Danos,1966-Ulrich,2010-Potylitsyn-Ryazanov}
\begin{equation}
\label{eq-decay}
W\sim\exp\left(-\frac{4\pi f}{c\beta\gamma} d\right),
\end{equation}
where $\gamma=(1-\beta^2)^{-1/2}$. 
This factor is proportional to energy density of the particle's near field in the frequency-domain, at distance $d$ in vacuum.
We verified that radiated energy in our numerical model obeys accurately proportionality (\ref{eq-decay}). In all results, we assume $d=70$~nm, which is experimentally achievable\cite{2022-Hausler-Seidling}. In this work we assume that 1st order radiation in the normal direction is at $f=410$~THz. At this frequency, the ``impact factor decay length'' $\frac{c\beta\gamma}{4\pi f}$ for radiated energy is 20~nm for 30~keV and 280~nm for 2~MeV electrons, and for $d=70$~nm factor (\ref{eq-decay}) is respectively equal to 0.031 and 0.78. This means that radiation from 30 keV electrons is very sensitive to the impact parameter.

\subsection{\label{sect-methods-decomposition}Decomposition into radiation orders}

We calculate the total radiated energy for one electron, $W$, in eV (Sect.~\ref{sect-30} and \ref{sect-2000}),  and its spectral density $\partial W/\partial f$
in eV/THz, separated into radiation orders $m$: $\partial W/\partial f=\sum_m \partial W_m/\partial f$
(Sect.~\ref{sect-spectra}). For a particular radiation order $m$,
radiated energy per unit frequency is\cite{2020-Szczepkowicz-Schachter-England}
\begin{equation}
    \frac{\partial W_m}{\partial f}=2\pi \frac{\partial W_m}{\partial\omega}
    =2\pi \int 4\cdot 2\pi\cdot\Re\left[\frac{1}{2}\mathbf{E}_m(\mathbf{r},\omega)\times \mathbf{H}_m^{*}(\mathbf{r},\omega)\right]\cdot d\mathbf{A}
\end{equation}
To calculate the fields $(\mathbf{E}_m, \mathbf{H}_m)$ of the $m$-th radiation order we first obtain from the simulation the full field inside the dielectric $(\mathbf{E}, \mathbf{H})$. For a given frequency $\omega$, this field can be expressed as a superposition of plane waves, both homogeneous and inhomogeneous\cite{1975-Jackson} (propagating and evanescent), which are Floquet-periodic along $z$ (see Fig.~\ref{fig-geometry}). For a $y$-invariant system, nonzero field components are $H_y, E_x$ and $E_y$. For example, $H_y$ takes the following generalized Fourier series expansion: 
\begin{equation}\label{eq-series}
    H_y(x,z;\omega)=
    \sum_{m=-\infty}^{\infty}H_y(x,z;\omega;m)=
    \sum_{m=-\infty}^{\infty}h_m(\omega)
    e^{-j (k-\Lambda m)z}
    \,
    e^{\sqrt{(k-\Lambda m)^2-(n \omega/c)^2}\,x},
\end{equation}
where $h_m(\omega)$ is the amplitude of $m$-th plane wave, $k=\omega/v$ is the Floquet wave number with phase velocity equal to the electron velocity, $\Lambda=2\pi/a$ is the spatial frequency of the grating, and $n$ is the refractive index of the grating support.
The series in Eq.~(\ref{eq-series}) consists of a small number of propagating plane waves with indices $m$ close to 0, and an infinite number of evanescent plane waves, where terms with high $|m|$ are negligible. 
If a term with some $m\neq0$ is propagating, it describes iSPR of order $m$. If the term $m=0$ is propagating, it describes CDR. The amplitude of the $m$-th radiation order is obtained from the full field using the standard Fourier series method:
\begin{equation}
    h_m(\omega)=
    \frac{1}{2a}
    \int_{0}^{a}  
    H_y(x,z;\omega)\,
    e^{j (k-\Lambda m)z}\,
    e^{-\sqrt{(k-\Lambda m)^2-(n \omega/c)^2}\,x}\,dz.
\end{equation}

\section{Results and discussion}

\subsection{\label{sect-30}Radiated energy from 30 keV electrons -- iSPR}

\begin{figure}
\includegraphics{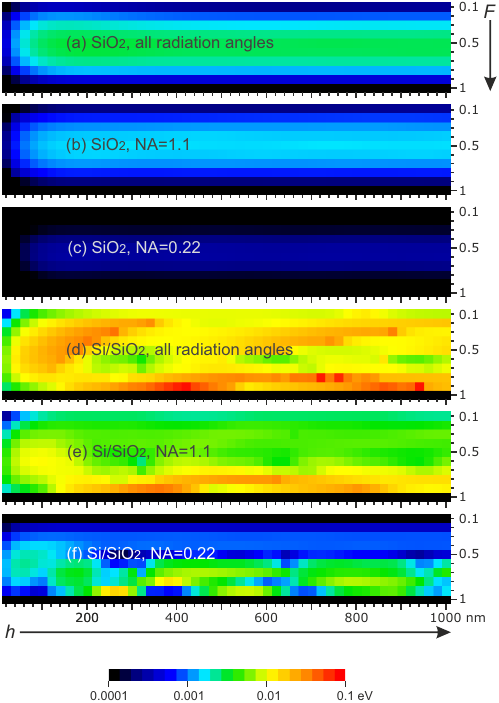}
\caption{Optimization of the geometry of SiO$_2$ (a--c) and Si/SiO$_2$ (d--f) for 30 keV electrons. In each of the insets (a--f), grating tooth height $h$ varies from 20~nm (left) to 1000~nm (right), while grating fill factor $F$ (defined in Fig.~\ref{fig-geometry}) varies from 0.1 (top) to 1.0 (bottom). The colors represent calculated total iSPR energy per electron.}
\label{fig-optimization-30}
\end{figure}

First we focus on the lower energy electrons: 30 keV. The corresponding $\beta=0.3284$ is below the threshold for generating Cherenkov Diffraction Radiation (CDR) in the silica substrate, and only positive Smith--Purcell radiation orders are observed ($m>0$). Figure \ref{fig-optimization-30} shows radiated energies for two material configurations: pure SiO$_2$ grating (SiO$_2$, insets a--c) or Si bars on SiO$_2$ substrate (Si/SiO$_2$, insets d--f). 
To optimize the grating geometry, we vary the grating tooth height: $h=20,40,60,\ldots,1000$~nm, and the grating fill factor: 
$F=0.1,0.2,0.3,\ldots,1.0$ (from 10\% to 100\% coverage, see Fig.~\ref{fig-geometry}).

For a pure SiO$_2$ grating, maximum radiated energy is obtained for grating fill factor $F=0.5$ and for any grating tooth height exceeding $h_0\approx160$~nm; this ``saturation height'' $h_0$ turns out to be smaller than the wavelengths for 1st order radiation ($m=1$, vacuum wavelengths 377--1079 nm, silica wavelengths 256--744 nm). For this type of grating, maximum radiated energy per electron is 0.0031 eV, as shown in Table~\ref{table-radiated-energy-30keV}.

{\renewcommand{\arraystretch}{1.2}
\begin{table}
  \caption{Calculated maximal radiated energy for 30 keV electrons.}
  \label{table-radiated-energy-30keV}
   \begin{minipage}{\textwidth}
   \begin{tabular}{|l|l|l|l|l|l|l|}
    \hline
    Radiation   & \multicolumn{6}{|l|}{Maximum radiated energy per electron [eV]}\\
    \cline{2-7}
    order $m$   &  \multicolumn{2}{|l|}{All radiation angles} & \multicolumn{2}{|l|}{Limited to NA=1.1} & \multicolumn{2}{|l|}{Limited to NA=0.22} \\
    \cline{2-7}
            & SiO$_2$ & Si/SiO$_2$ & SiO$_2$ & Si/SiO$_2$ & SiO$_2$ & Si/SiO$_2$ \\
    \hline
 
    $+1$  & 0.0030&0.0900  &  0.0015&0.0402  &  0.0003&0.0165  \\
    $+2$  & 0.0001&0.0024  &  0.0001&0.0012  &  0.0000&0.0001  \\
    $+3$  & 0.0000&0.0001  &  0.0000&0.0000  &  0.0000&0.0000  \\
    $+4$  & 0.0000&0.0000  &  0.0000&0.0000  &  0.0000&0.0000  \\
    All orders\footnote{Maximal energies for individual orders are achieved each for a different geometry, therefore maximal results should not be summed over $m$.} & 0.0031&0.0904  &  0.0016&0.0404   & 0.0003&0.0165 \\
\hline
  \end{tabular}
   \end{minipage}
\end{table}}

For the electron energy that is discussed in this section (30 keV -- below the Cherenkov threshold), one can obtain higher radiated energies for another type of grating: Si/SiO$_2$, as evidenced in Fig.~\ref{fig-optimization-30} and Table~\ref{table-radiated-energy-30keV}. The higher scattered energy may be explained by the higher index of refraction of the scatterers -- the Si grating teeth, compared to a pure SiO$_2$ grating. 
For the Si/SiO$_2$ grating, the maximum radiated energy per electron is 0.0904 eV, but the question of optimal geometry is more complex -- the maxima of radiated energy are sensitive to variations of the grating geometry. This can be explained by the fact that Si bars surrounded by lower index regions act as resonators, and the grating eigenmodes interfere with SPR and iSPR\cite{2018-Yang-Massuda, 2021-Hu-Hu, 2022-Sirenko-Poyedinchuk, 2023-Chen-Mao, 2023-Yang-Roques-Carmes}. The analysis of this effect is beyond the scope of this work. We just note that the analysis of the photonic band diagram alone is not sufficient, as it fails to explain the differences between maxima for SPR and iSPR, and also between different radiation orders.

For the Si/SiO$_2$ grating, the optimal geometry depends on the numerical aperture of the experimental setup, as seen in insets (d--f) in Fig.~\ref{fig-optimization-30}. If radiation at all angles could be collected, a good choice for the geometry would be $F=0.9$, $h=420$~nm (0.090 eV/electron) or $F=0.8$, $h=720$~nm (0.082 eV/electron). If radiation is to be guided with a SiO$_2$ fiber with plastic coating removed (NA = 1.1), optimal grating parameters are $F=0.9$, $h=360$~nm (0.040 eV/electron) or $F=0.8$, $h=620$~nm (0.033 eV/electron). For the off-the-shelf fiber with NA=0.22, good grating parameters are $F=0.9$, $h=260$~nm (0.016 eV/electron) or $F=0.8$, $h=380$~nm (0.014 eV/electron). This radiation efficiency is considerably limited by the small value of the exponential factor in Eq.~(\ref{eq-decay}) for the assumed energy 30~keV and impact parameter $d=70$~nm (Sect.~\ref{sect-impact-par}); note that for grazing electrons ($d\approx0$~nm) the energy would be 30 times higher.

Radiation efficiency described above is expressed in electronvolts per electron. Alternatively, the number of emitted photons per electron can be given, defined as 
\begin{equation}
    N=\int_{200~\mathrm{THz}}^{1200~\mathrm{THz}} \frac{1}{hf} \frac{\partial W}{\partial f} df
\end{equation}
Maximal radiation for SiO$_2$ grating, 0.0031 eV/electron, corresponds to 0.0022 photons/electron, while for Si/SiO$_2$ grating, maximal radiation is 0.090 eV/electron, or 0.072 photons/electron. 

For both types of gratings, pure SiO$_2$ and Si/SiO$_2$, radiation is zero for grating fill factor $F=1$, corresponding to a flat uniform surface, because radiation orders $m\neq 0$ require periodic changes of the refraction index along the electron path. The situation is different if the electron velocity is increased above the Cherenkov threshold, where radiation order $m=0$ is present also for a flat uniform surface, as shown in the next section.

\subsection{\label{sect-2000}Radiated energy from 2 MeV electrons -- CDR\&iSPR}

\begin{figure}
\includegraphics{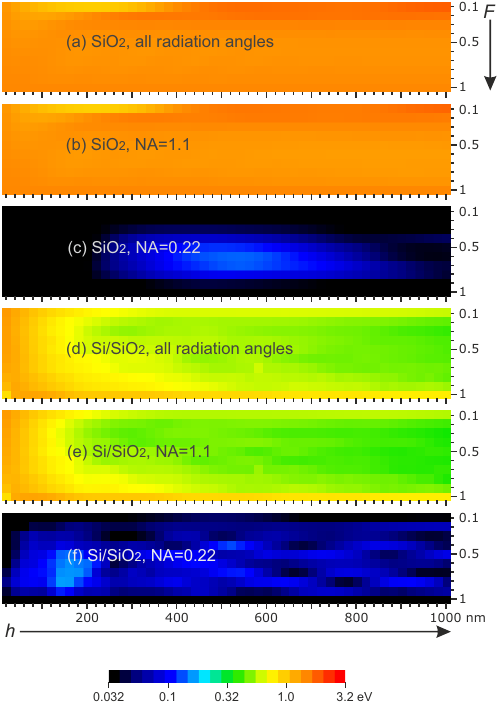}
\caption{Optimization of geometry as in Fig.~\ref{fig-optimization-30}, but for higher energy electrons: 2 MeV. In the present case, the colors represent the sum of iSPR and CDR energy. Note that the color scale is different than in Fig.~\ref{fig-optimization-30}.}
\label{fig-optimization-2000}
\end{figure}

Next we present results for higher energy electrons: 2 MeV. The corresponding $\beta=0.9791$ is above the threshold for generating Cherenkov Diffraction Radiation (CDR) in the silica substrate,
so, in addition to iSPR ($m\neq0$), CDR is emitted ($m=0$). Figure \ref{fig-optimization-2000} shows total radiated energies for different grating geometries $(h,F)$ and two material configurations: SiO$_2$ (a--c), and Si/SiO$_2$ (d--f). The corresponding energy values are shown in Table~\ref{table-radiated-energy-2MeV}.

{\renewcommand{\arraystretch}{1.2}
\begin{table}
  \caption{Calculated maximal radiated energy for 2 MeV electrons.}
  \label{table-radiated-energy-2MeV}
   \begin{minipage}{\textwidth}
   \begin{tabular}{|l|l|l|l|l|l|l|}
    \hline
    Radiation   & \multicolumn{6}{|l|}{Maximum radiated energy per electron [eV]}\\
    \cline{2-7}
    order $m$   &  \multicolumn{2}{|l|}{All radiation angles} & \multicolumn{2}{|l|}{Limited to NA=1.1} & \multicolumn{2}{|l|}{Limited to NA=0.22} \\
    \cline{2-7}
            & SiO$_2$ & Si/SiO$_2$ & SiO$_2$ & Si/SiO$_2$ & SiO$_2$ & Si/SiO$_2$ \\
    \hline
 
    $\phantom{-}0$ (CDR) & 1.75&1.37  &  1.75&1.37  &  0&0  \\
    $+1$       &  0.36&0.32  &  0.36&0.32  &  0.05&0.14  \\
    $+2$       &  0.13&0.15  &  0.12&0.12  &  0.07&0.04  \\
    $+3$       &  0.08&0.08  &  0.07&0.05  &  0.03&0.02  \\
    $+4$       &  0.07&0.06  &  0.06&0.05  &  0&0  \\
    $-1,+5,+6$ &  0.10&0.09 & ND\footnote{ND = not determined.} & ND  &  ND & ND \\
    All orders\footnote{Maximal energies for individual orders are achieved each for a different geometry, therefore maximal results should not be summed over $m$.} &  1.97&1.40  &  $\sim$1.89&$\sim$1.39  &  $\sim$0.13&$\sim$0.16 \\
\hline
  \end{tabular}
   \end{minipage}
\end{table}}

Compared with 30~keV results, 2 MeV electrons at the same impact parameter produce approximately an order of magnitude more radiated energy, reaching 1.97~eV (0.87 photons) per electron for SiO$_2$ grating with high narrow teeth ($F=0.1$, $h=1000$~nm). This radiation increase with electron energy increase is partially due to higher ``energy decay length'' as described in Sect.~\ref{sect-impact-par}. Another factor is the occurrence of the $m=0$ radiation order: CDR, which is typically several times stronger than iSPR. However CDR is not captured in the low numerical aperture setup, NA = 0.22, which results in relatively smaller total energy for this aperture. For high energy electrons, the dependence of total radiated energy on geometry is weaker due to dominance of CDR, and, as expected, CDR also occurs for the uncorrugated surface ($F=1$).

An unexpected result for 2~MeV electrons is that the lower index grating, pure SiO$_2$, yields higher radiation than Si/SiO$_2$. The reason for this is unclear and needs further study.

\subsection{\label{sect-spectra}The spectra for different radiation modes}

One of the distinct features of this work is that the numerically calculated iSPR and CDR radiation is decomposed into radiation orders. This needs to be done in order to analyze the angles captured within different numerical apertures, but may also be of general theoretical interest, inspiring for future investigations. 

\begin{figure}
\includegraphics{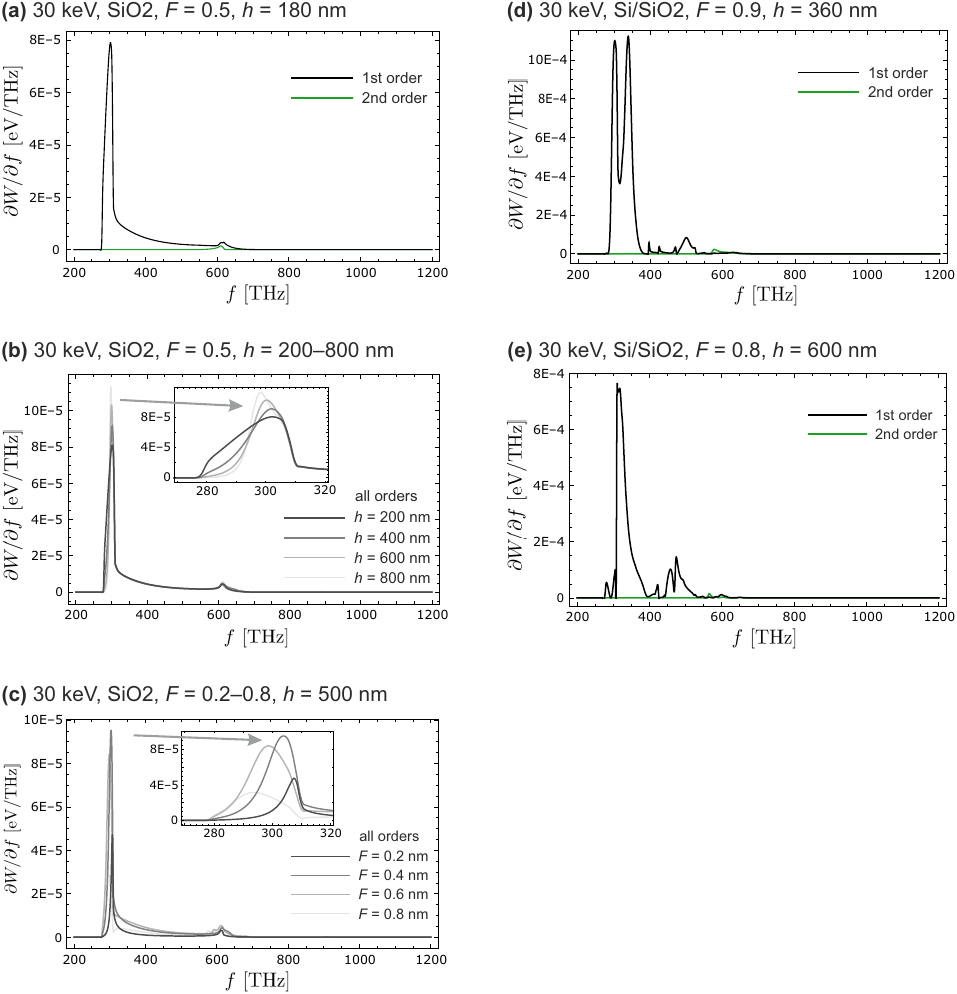}
\caption{Example iSPR spectra for 30 kV electrons.
(a--c) SiO$_2$ grating, (d, e) Si/SiO$_2$ grating.}
\label{fig-spectra-30}
\end{figure}

First, in Fig.~\ref{fig-spectra-30}, we present example decomposed spectra for 30~keV electrons. At this electron energy the radiation is dominated by the $m=1$ order. Small contribution from $m=2$ radiation order is shown in green, while contribution of higher orders is negligible. For SiO$_2$ grating, first order radiation is maximal around 300 THz (Fig.~\ref{fig-spectra-30}(a)).
The frequency of this radiation peak is not significantly dependent on grating tooth height (Fig.~\ref{fig-spectra-30}(b)) and shows a small drift with grating fill factor (Fig.~\ref{fig-spectra-30}(c)): $F=0.2{-}0.8$ $\Rightarrow$ $f=308{-}293$~THz, corresponding to emission angles $\theta=134^{\circ}{-}147^{\circ}$. The frequency of this peak is not far from the threshold of 1st order emission (278 THz, see table~\ref{table-f-ranges-30}), which corresponds to $\theta=180^{\circ}$ (radiation in the backward direction). A similar radiation peak was reported for a pure Si grating at the same electron energy, but its nature is yet unexplained by existing theoretical models\cite{2021-Konakhovych-Sniezek,2022-Konakhovych-Sniezek}. As shown in Fig.~\ref{fig-spectra-30}(d,e), the spectra for Si/SiO$_2$ gratings are more complex, which is a result of a complex photonic band structure of the resonant Si/SiO$_2$ system (see also Sect.~\ref{sect-30}). 

\begin{figure}
\includegraphics{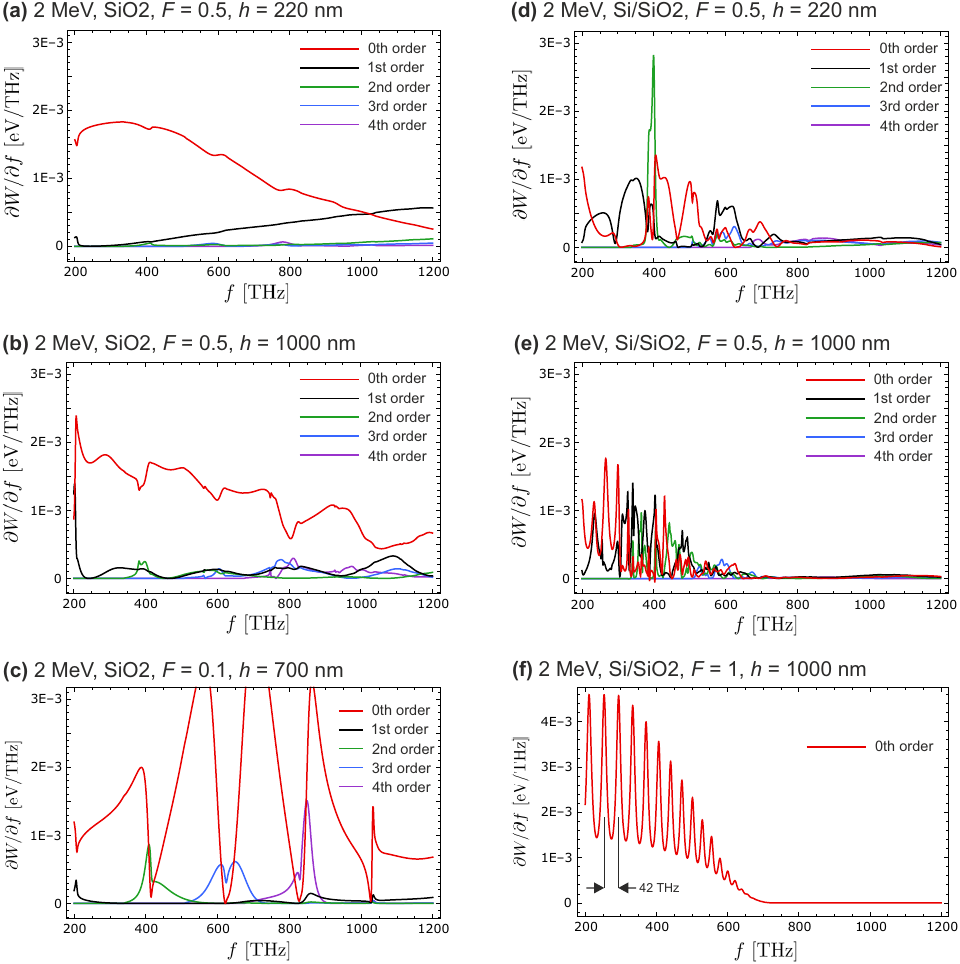}
\caption{Example iSPR spectra for 2 MeV electrons.
(a--c) SiO$_2$ grating, (d--f) Si/SiO$_2$ grating.}
\label{fig-spectra-2000}
\end{figure}

Figure~\ref{fig-spectra-2000} shows example decomposed spectra for higher energy (2~MeV) electrons. The spectra are now dominated by the $m=0$ radiation order, CDR (red curve). These high electron energy spectra contain higher frequency radiation (above 1200 THz) compared to 30 keV electrons (where radiation was negligible above 700 THz). Also, there is significant contribution from higher radiation orders. For example, Fig.~\ref{fig-spectra-2000}(d) shows that $m=2$ may dominate over all other orders in a narrow frequency range, while (b) and (c) show noticeable contribution from the 4th radiation order. Cherenkov diffraction radiation shows a highly resonant behavior\cite{2022-Karataev-Naumenko}, in contrast to CDR near a planar surface of a uniform material\cite{1966-Ulrich,2021-Konakhovych-Sniezek,2022-Konakhovych-Sniezek}. For the SiO$_2$ grating, the minima in the CDR spectrum are close to the maxima of the other radiation orders (a--c), which might be connected with conservation of energy, while the resonant Si/SiO$_2$ gratings (d,e) exhibit extremely complex spectra.

Finally, Fig.~\ref{fig-spectra-2000}(f) shows a very regular comb-like spectrum for a special case where the Si layer reaches fill factor $F=1$ and forms a 1000-nm uniform film atop a SiO$_2$ substrate. Such silicon layer forms a Fabry--Perot resonator. Note however that in our case the resonator is excited by an \emph{evanescent} wave, in contrast to the well-known textbook traveling wave case. The free spectral range $(c/n)/(2h)$ near 250~THz is 42.5~THz, which is consistent with the oscillation period from the numerical simulation. However, near 550~THz $(c/n)/(2h)=36.5$~THz, while the numerical oscillation period is 25~THz -- 30\% lower then expected from the simple perpendicular propagating wave model.  

\subsection{Inverse design} 

Recently, several advanced multiparameter optimization schemes for photonic design were proposed, including photonic inverse-design\cite{2018-Molesky-Lin, ceviche-workshop}. Within this research project, we also tried to design better gratings for iSPR using inverse design, beyond the simple rectangular gratings described above. Our inverse design work was limited to 30 keV electrons, pure SiO2 grating, constant refractive index, and NA=0.22. We found that it is possible to find dielectric structures that enhance iSPR by an order of magnitude, but these structures contain hollow isolated vacuum regions (the dielectric is not \emph{simply connected} in the topological sense). Such structures are fabricable by etching or milling when the invariant direction is perpendicular to the wafer\cite{2022-Hausler-Seidling}. However in our configuration for fiber-coupled iSPR, the invariant direction $y$ is tangent to the wafer (Fig.~\ref{fig-geometry}), and isolated vacuum regions are not fabricable. 
In another inverse-design attempt, we limited the structure space to structures that can be constructed by etching a Si/SiO2 wafer in the normal direction. Within this structure space, we were not able to find structures which radiate better than optimized simple rectangular gratings.

\section{Conclusion}

Electrons travelling in vacuum in proximity of a dielectric grating generate light which propagates inside the dielectric. 
Internal Smith--Purcell Radiation (iSPR) and Cherenkov Diffraction Radiation (CDR) are different terms of the plane wave expansion of the total radiation field inside the dielectric.

We proposed a setup for measuring visible and near-visible iSPR and CDR, consisting of a dielectric grating (SiO$_2$ or Si/SiO$_2$) attached directly to a silica multimode optical fiber, and performed quantitative numerical analysis of its radiation efficiency, both in terms of total radiated energy per electron and in terms of spectral distribution of different radiation orders $m$, where $m=0,1,2,3,4$. The total iSPR and CDR radiated energy, assuming grating length of 200~$\mu$m, reaches 2~eV per electron for electron velocities above the Cherenkov threshold and rectangular SiO$_2$ grating with high narrow teeth. For lower velocities, radiation efficiency is limited by small range of the electron's near field. Above the Cherenkov threshold, in most cases the energy of CDR is several times higher than the energy of iSPR, but for some geometries and frequency ranges first or second order iSPR may have the highest energy spectral density. The spectrum of CDR is highly resonant, with local minima for frequencies where maxima of the other radiation orders occur. The CDR spectrum takes a frequency comb-like shape for a uniform layer of Si on SiO2 substrate, due to a Fabry-Perot effect occurring for the evanescent field of the moving electron.
Both iSPR and CDR radiation spectra are sensitive to the energy of the electron beam, so the proposed setup could become a prototype of a non-invasive particle beam monitor for both conventional and miniaturised laser particle accelerators.

\begin{acknowledgement}

A.S. thanks Zhexin Zhao, Alexey A.\ Tishchenko, and Urs H\"{a}usler for discussions.

\noindent \textbf{Funding:}\\
Gordon and Betty Moore Foundation (GBMF4744);

\end{acknowledgement}

\bibliography{article-si-sio2}

\end{document}